\documentclass[10pt,onecolumn]{revtex4}
\usepackage{graphicx} 
\usepackage{dcolumn}
\usepackage{bm}
\usepackage{hyperref}
\pdfoutput=1
\usepackage{amssymb} 
\usepackage{amsmath}
\usepackage{color}

\newcommand{\BE}{\begin{equation}}
\newcommand{\EE}{\end{equation}}
\newcommand{\BA}{\begin{eqnarray}}
\newcommand{\EA}{\end{eqnarray}}
\newcommand{\bx}{{\bf x}}

\newcommand{\bq}{{\bf q}}

\newcommand{\ba}{{\bf a}}

\newcommand{\bxi}{{\boldsymbol{\xi}}}

\newcommand{\bhn}{{\boldsymbol{\hat n}}}
\newcommand{\bP}{{\boldsymbol{P}}}
\newcommand{\bfr}{{\boldsymbol{F}}_\rho}

\begin{document}

\title{Active cluster crystals}

\author{Jean-Baptiste Delfau}
\affiliation{IFISC (CSIC-UIB), Instituto de F\'{\i}sica
Interdisciplinar y Sistemas Complejos, Campus Universitat de
les Illes Balears, E-07122 Palma de Mallorca, Spain}

\author{Crist\'{o}bal L\'opez}
\affiliation{IFISC (CSIC-UIB), Instituto de F\'{\i}sica
Interdisciplinar y Sistemas Complejos, Campus Universitat de
les Illes Balears, E-07122 Palma de Mallorca, Spain}

\author{Emilio Hern\'andez-Garc\'{\i}a}
\affiliation{IFISC (CSIC-UIB), Instituto de F\'{\i}sica
Interdisciplinar y Sistemas Complejos, Campus Universitat de
les Illes Balears, E-07122 Palma de Mallorca, Spain}

\date{\today}

\begin{abstract}
We study the appearance and properties of cluster crystals
(solids in which the unit cell is occupied by a cluster of
particles) in a two-dimensional system of self-propelled active
Brownian particles with repulsive interactions. Self-propulsion
deforms the clusters by depleting particle density inside, and
for large speeds it melts the crystal. Continuous field
descriptions at several levels of approximation allow to
identify the relevant physical mechanisms.
\end{abstract}


\maketitle

\section{Introduction}

The collective behavior of self-propelled particles is a
fascinating topic both for its numerous applications
and its intrinsic theoretical interest
\cite{Ramaswamy2010,Cates2010, Marchetti2013,Romanczuk2012}.
Many studies focused on the formation of clusters reporting two
main different cases: for ``active crystals'', self-propulsion
leads to a modification of the properties of a pre-existing
crystal that is usually induced by long-range attractive and
short-range repulsive interactions
\cite{Mani2015,Menzel2014,Menzel2013}. On the
other hand, for Mobility-Induced Phase Separation (MIPS), the
system separates into two fluid phases of different densities.
Clusters of the densest phase form by a purely non-equilibrium
mechanism induced by the persistence of the motion of the
particles, the nature of their interactions being not as
crucial (this phenomenon was observed for various repulsive and
attractive forces \cite{Cates2015,Mognetti2013,Redner2013}).
Methodologically, both mechanisms can be studied by considering
an effective density-dependent velocity replacing the two-body
interacting potential
\cite{Tailleur2008,Fily2012,Farrell2012,Cates2013,Stenhammar2013,Bialke2013,Fily2014}.

We analyze in this paper a different case of cluster formation
with active objects (some additional models and
experiments on clustering of self-propelled particles maybe
found, for example, in
\cite{Buttinoni2013,Redner2013,Theurkauff2012}). It is the
non-equilibrium counterpart of the so-called cluster crystals
\cite{Klein1994,Likos2001,Mladek2006,Likos2007,Mladek2008,Coslovich2013},
which appear in equilibrium systems interacting with soft-core
repulsive potentials, and are solid-like structures where the
unit cell is occupied by a closely packed cluster of particles.
Here, clustering appears under a repulsive potential so that it
is not a consequence of purely local effects but rather
involves a global minimization of energy by intercluster
interactions. Namely, for sufficiently soft repulsive
potentials, repulsion from the neighboring clusters can exceed
the intracluster repulsion and lead to an effective particle
confinement in the cluster. For the active-particle case
discussed here, this mechanism is very different from  the
other cases of active crystals and it is unlikely that previous
local arguments derived for active crystals or MIPS can
describe it.

Similarly to previous works \cite{Menzel2014,Mani2015}, we
consider the simplest extension of an equilibrium system of
repulsive Brownian particles in two dimensions by providing
them with an internal degree-of-freedom, the orientation of a
self-propulsion speed. We will address the following questions:
Do {\it active} cluster crystals (ACC) with
only repulsive interactions exist? Is the structure of the
clusters modified by activity? Can we find continuum equations
describing this active system? The answer to
the first question has been partially given by observations of
cluster crystals in \cite{Menzel2012,Menzel2013laning}, for
example. But in these references the focus is in
deformable-body and alignment interactions, so that the
specific role of repulsive interactions needs further
clarification.

To do this and answer the remaining questions we first present
numerical simulations of self-propelled soft
repulsive particles showing that ACC can be observed but that
an increasing self-propulsion eventually leads to their
destruction. For small diffusion, {\it empty clusters} are
found as the particles tend to accumulate on their edges. Then
we provide a continuum field description of this interacting
particle system and analyze some of its predictions.

\section{Numerical results and pattern formation}

\begin{figure}
\centerline{\includegraphics[height=14 cm]{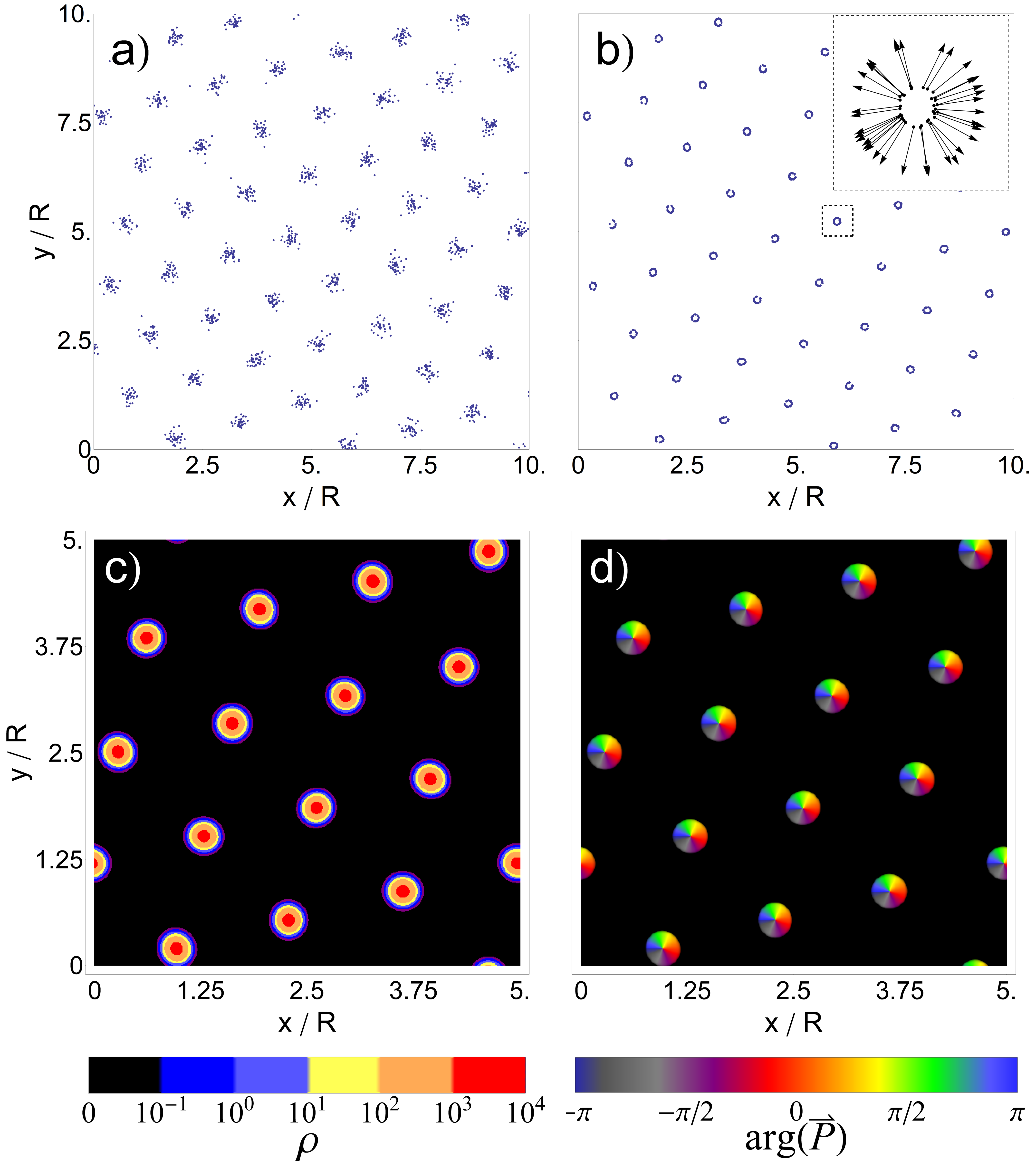}}
\caption{a) and b) Particle simulations: snapshots of the positions of
the particles in the steady-state for $\bar\rho = 2000$, $L=1$, $D_r = 0.1$ and
a GEM-3 potential with $R=0.1$ and $\epsilon=0.0333$. a) $D = 3 \times 10^{-2}$,
$U_0 = 1.0$. b) $D = 10^{-4}$ , $U_0 = 3.0$. The top-right inset is a zoom of the
boxed cluster showing the orientation vector $\bhn$ of the particles inside it. c) and d) Steady-states
obtained by numerical integration of Eqs.~(\ref{DK_ABP_fin}). Same parameters as in panel a)
except $L=0.5$. c) Local density $\rho$ . d) Orientation angle $\arg\boldsymbol{P}$ of the
polarization field, encoded in colors. The modulus $|\boldsymbol{P}|$ is encoded in the
opacity, so that the black areas correspond to very small polarizations.}
\label{fig:snap_ldensity}
\end{figure}

We start with the Langevin equations of a two-dimensional
overdamped system of active Brownian particles interacting in pairs via
a potential $V(\bx)$:
\BA
\label{ABP_matrix}
\dot \bx_i(t) &=& U_0 \bhn(\theta_i(t)) -
\sum_{j\ne
i} \nabla V(\bx_i-\bx_j) + \sqrt{2D} \bxi_i(t), \nonumber \\
\dot\theta_i(t) &=& \sqrt{2D_r} \xi_i^r(t) \ .
\EA
$\left(x_i,y_i \right) = \bx_i$ and $\theta_i$ are respectively
the position and orientation of the particle $i$, $U_0\bhn$ is
the self-propelling velocity, of constant modulus $U_0$ and
direction given by the unit vector
$\bhn(\theta_i)=(\cos\theta_i,\sin\theta_i)$. The particles are subjected to Gaussian
translational and rotational noises, $\bxi_i=(\xi_i^x,\xi_i^y)$
and $\xi^r_i$ respectively, satisfying
$\left<\xi_i^\alpha\right> = 0$, and
$\left<\xi^\alpha_i(t)\xi^\beta_j(t')\right> =
\delta_{ij}\delta_{\alpha\beta}\delta(t-t')$, for
$\alpha,\beta=x,y,r$. $D$ and $D_r$ are the translation and
rotational diffusion coefficients. If they
originate from the same thermal bath they are related to
temperature by fluctuation-dissipation relationships. But under
general non-equilibrium conditions they can have different
origins and then we will assume here that they are independent
parameters.

Cluster crystals appear at equilibrium ($U_0=0$) under
repulsive soft-core potentials which have negative Fourier
components \cite{Likos2007,Delfau2016}. A convenient class of
such potentials is the generalized exponential model of
exponent $\alpha$ (GEM-$\alpha$): $V(\bx) = \epsilon
\exp\left(-\left|\bx/{R}\right|^\alpha\right)$, $\epsilon>0$, and $R$ is the interaction range.
The Fourier transform $\tilde V(q)$ of GEM-$\alpha$ potentials
with $\alpha>2$ takes both positive and negative values, so
that cluster crystals would appear in that case for
sufficiently small $D$ \cite{Likos2007,Delfau2016} and/or large average densities $\bar{\rho}$. Here we
consider $\alpha=3$.

Besides the global parameters $N$ and $L$,
which fix the mean density $\bar\rho=N/L$, our model contains
five parameters: $D$, $D_r$, $U_0$, and the two parameters in
the potential ($\epsilon$ and $R$). By proper choosing of space
and time units, all dependence in these five parameters gets
condensed in a set of three dimensionless quantities, for which
one of the possible choices is $U_0/\sqrt{D_r D}$, $\epsilon
D_r/U_0^2$, and $D_r R/U_0$. The first parameter compares the
strength of self-propulsion with that of translational
diffusion, the second the strength of the interaction forces
with self-propulsion, and the third the persistence length with
the interaction length. Unless otherwise stated, in the
following we will fix the parameters in the potential (to
$\epsilon=0.0333$ and $R=0.1$) and $D_r=0.1$ (this last one
being equivalent to fixing the units of time), and describe the
behavior of the system by varying $U_0$ and $D$. Other ways to
explore the transitions occurring in the system are possible,
and they can be easily related to the one described here simply
by looking at the dimensionless quantities stated above. For
example, in the dimensionless parameters the dependence on
$D_r$ is reciprocal to that on $U_0$, meaning that the changes
in behavior described below when increasing $U_0$ at $D_r$
fixed will also be observed when decreasing $D_r$ at $U_0$
fixed.

\begin{figure}
\centerline{\includegraphics[width=0.4\columnwidth]{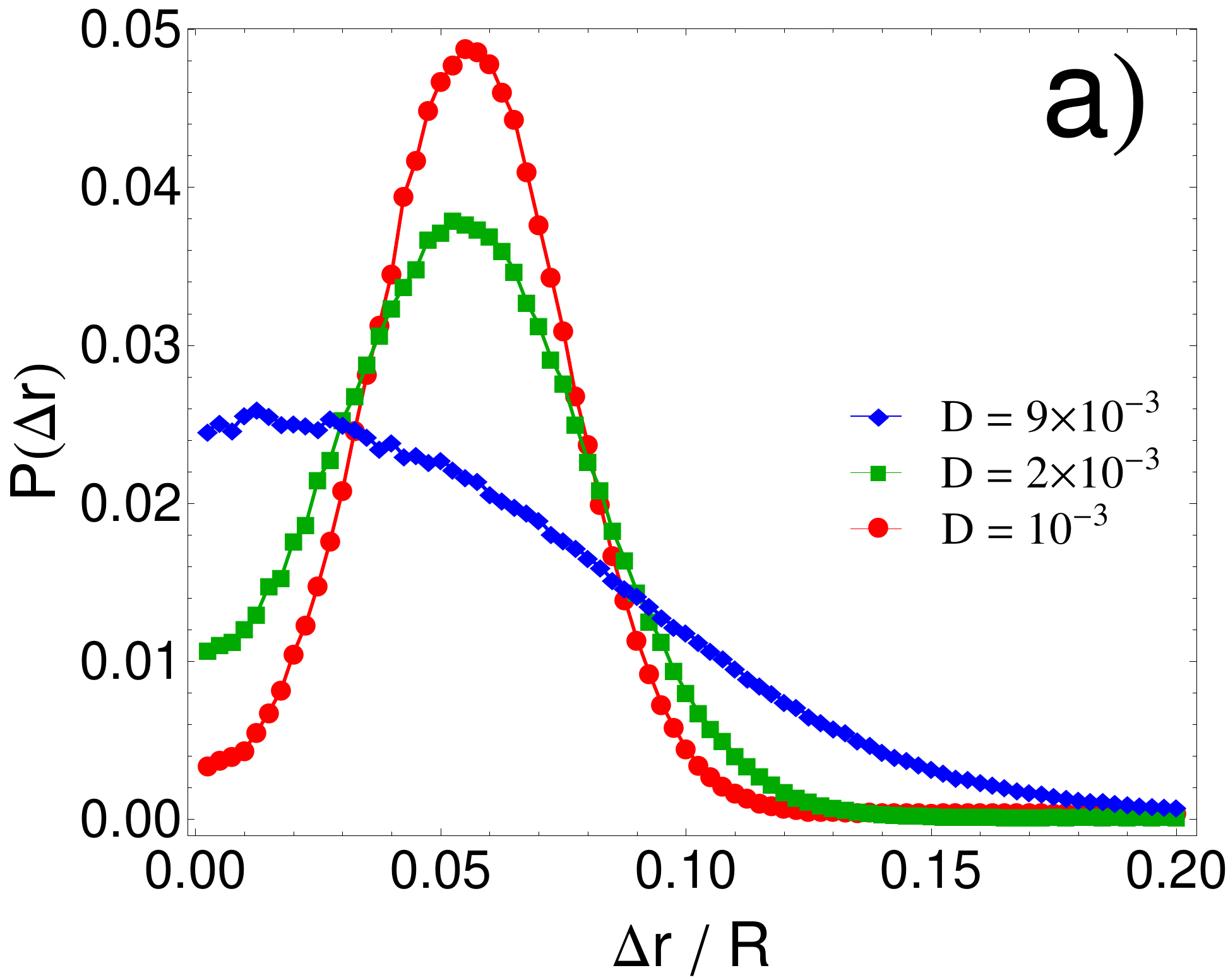}
\hspace{0.5cm}
\includegraphics[width=0.4\columnwidth]{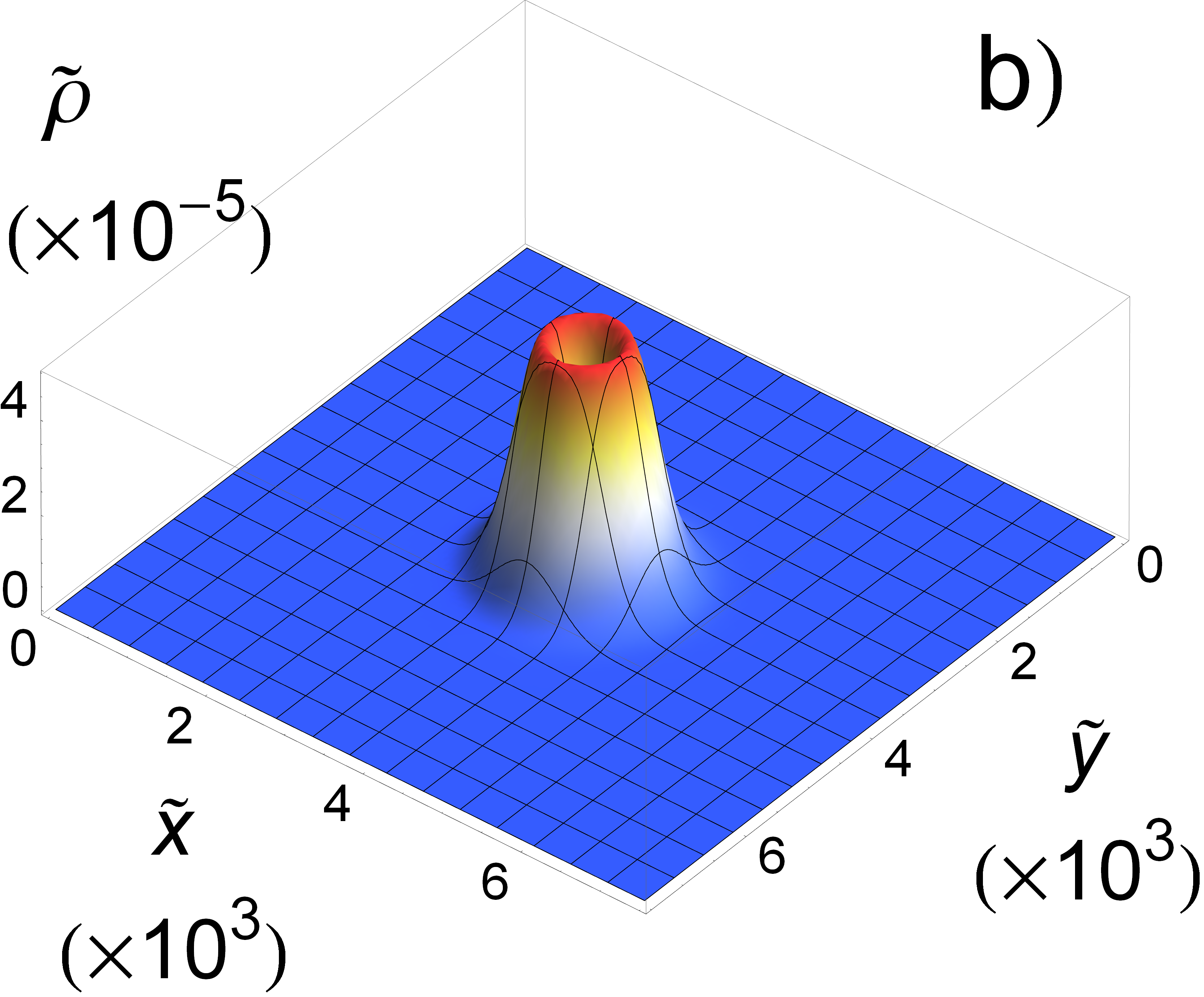}}
\caption{a) Probability of finding a particle at a distance
$[\Delta r,\Delta r + dr]$ from the center of its cluster, normalized by $2\pi \Delta r dr$ to give an estimation of the
particle density. Same parameters as in figure \ref{fig:snap_ldensity} a) and b)
except $U_0 = 2.5$, and $D = 10^{-3}$ (red disks), $2 \times 10^{-3}$ (green squares),
and $9 \times 10^{-3}$ (blue diamonds). b) Steady-state solution of Eqs.(\ref{DK_ABP_fin}) showing a 2D empty cluster. The interaction force $\boldsymbol{F}_{\rho}$ of Eqs.(\ref{DK_ABP_fin}) was replaced by an effective confinement force $\boldsymbol{F}_{eff} = \left( \tilde{a} \sin(2 \pi (\tilde{x}/\tilde{L}-0.5)) , \tilde{a}\sin(2 \pi (\tilde{y}/\tilde{L}-0.5)) \right) $. The parameters have been nondimensionalized by $D_r$ and $U_0$ so that $\tilde{a} = a/U_0=2.5$, $\tilde{L} = L D_r/U_0=7.5 \times 10^{-3}$, $\tilde{D}=D D_r / U_0^2 = 1.88 \times 10^{-3}$. The integral of $\tilde{\rho}$ is equal to 1.
} \label{fig:clust_structure}
\end{figure}

We integrate numerically (\ref{ABP_matrix})
with the Euler-Maruyama method \cite{Toral2014}. For small
$U_0$ we observe a transition, when increasing the average
density $\bar{\rho}$ or decreasing the diffusion coefficient
$D$, from a homogeneous distribution of particles to a
statistically steady hexagonal crystal of clusters (see
figure~\ref{fig:snap_ldensity} a,b), similar to the behavior
observed for passive particles ($U_0=0$). If $U_0$ and $D$ are
small enough, there is no particle exchange between clusters.
If $U_0$ is further increased, the cluster crystal remains but
jumps between clusters eventually occur. For high values of
$U_0$ clusters are destroyed and the steady-state becomes
statistically homogeneous (see Table~\ref{SM:table1} for
precise values). Contrary to the passive case for which the
distribution of particles inside a cluster is Gaussian to a
good approximation\cite{Delfau2016}, active particles tend to
stay close to the edges. This is particularly obvious for small
$D$ and high $U_0$. In that case, the clusters have a ring
shape with their centers empty (see
figure~\ref{fig:snap_ldensity} b). We confirmed this by
computing the distribution of $\Delta r$, the relative distance
between the position of a particle and the center of its
cluster. Figure~\ref{fig:clust_structure} a) shows that when
$U_0$ is large and $D$ small, there is a clear depletion close
to $\Delta r = 0$ while the distributions look Gaussian for
higher $D$ (remember than the relevant
dimensionless parameter is $U_0/\sqrt{D_r D}$, so that simple
Gaussian clusters are also obtained for large enough $D_r$).
A similar cluster-center depletion~\cite{Nash2010, Menzel2015, Menzel2016} 
has been observed in active systems in which aggregation occurs by 
confinement in an external potential. Our studies in 
Sect.~\ref{2modes_eq} below will show that indeed both phenomena are related. 
They arise from a purely non-equilibrium effect caused by 
the persistence of the particle velocity which
disappears for small $U_0$ or large $D_r$. Another consequence
of persistence is that the orientation $\bhn$ of the particles
inside a cluster is clearly radial, pointing to the exterior
(inset of figure~\ref{fig:snap_ldensity} b). For larger $D$
like in figure~\ref{fig:snap_ldensity} a), this radial
orientation of velocity vectors remains predominant but not as
strong. From the definition of the model, and
as we will show later on, the equilibrium Brownian non-active
dynamics, for which clusters are no longer empty, is achieved
for large $D_r$.

\section{Theoretical description: active Dean-Kawasaki equation}

\subsection{Derivation and stability analysis}

To have some analytical handle on the phenomena above we derive
now the equivalent of a Dean-Kawasaki (DK) equation
\cite{Kawasaki1994,Dean1996,Delfau2016} for active particles.
Note that a continuum description equivalent to
the particle system should include a noise term arising from
the random discrete particle dynamics. However, as in
\cite{Delfau2016} we average out noise terms to obtain a
deterministic version of the DK equation. To do so, we start
from (\ref{ABP_matrix}). The function $\hat\rho(\bx,\theta,t) =
\sum_i^N\delta( \bx-\bx_i(t))\delta(\theta-\theta_i(t))$ is an
It\={o} process so that we can apply the It\={o}
formula\cite{Oksendal}. Using integration by parts we obtain
the following deterministic equation for the average
$\rho(\bx,\theta,t)\equiv \left<\hat\rho(\bx,\theta,t)\right>$,
which is the expected value of the density of particles at
location $\bx$ and with orientation $\theta$:
\BA
\partial_t \rho(\bx,\theta,t) = \nabla \cdot \left(
\rho(\bx,\theta,t) \int d\bx' \nabla V(\bx-\bx')
\rho(\bx',t)  \,  \right) && \nonumber \\
+ D \nabla^2 \rho(\bx,\theta,t) + D_r
\partial^2_\theta \rho(\bx,\theta,t) - U_0 \bhn(\theta)\cdot
\nabla \rho(\bx,\theta,t) . && \label{DK_ABP}
\EA
$\rho(\bx',t)=\int_0^{2\pi} d\theta \rho(\bx',\theta,t)$ is the
total particle density at point $\bx'$ and ${\bf \hat
n}(\theta)=(\cos\theta,\sin\theta)$ is the unit vector in the
$\theta$ direction. Equations similar to (\ref{DK_ABP})
have already been derived by other means to describe MIPS
\cite{Bialke2013,Speck2015} but here we do not approximate the
non-local interaction term in terms of a local
density-dependent velocity, and instead we have introduced a
mean-field approximation:
$\left<\hat\rho(\bx,\theta)\hat\rho(\bx',\theta')\right>\approx
\rho(\bx,\theta)\rho(\bx',\theta')$. For passive particles
interacting with soft potentials, this is usually a good
approximation because of the large number of particles within
the interaction range \cite{Delfau2016}.

An equivalent representation is obtained by introducing the
angular Fourier transform: $\rho_n(\bx,t) \equiv \int_0^{2\pi}
\rho(\bx,\theta,t) e^{i n \theta} d\theta$ for $n=0,\pm 1,\pm
2, ...$. Note that $\rho_0(\bx,t)=\rho(\bx,t)$. From
(\ref{DK_ABP}) we find a coupled set of equations for the
angular modes $\rho_n$:
\BA
\label{DK_ABP_mode}
\partial_t \rho_n(\bx,t) &= -\nabla \cdot \left( \bfr(\bx,t)\rho_n(\bx,t)
\right)
+ D \nabla^2 \rho_n(\bx,t) - D_r n^2 \rho_n(\bx,t) \\
&- U_0 \left( \frac{\partial_x \rho_{n+1}(\bx,t) +
\partial_x \rho_{n-1}(\bx,t)}{2} \right.
 + \left.\frac{\partial_y \rho_{n+1}(\bx,t) -
\partial_y \rho_{n-1}(\bx,t)}{2 i} \right)\ \nonumber ,
\EA
where
\BE \bfr(\bx,t)\equiv -\int \nabla V(\bx-\bx')
\rho(\bx',t) d\bx'
\label{fadvective}
\EE
is the force induced at point $\bx$ by the particle
interactions.

The homogeneous and isotropic state given by
$\rho(\bx,\theta,t)=\bar\rho$, or
$\rho_n(\bx,t)=\bar\rho\delta_{n0}$, is a steady solution of
Eqs. (\ref{DK_ABP}) or (\ref{DK_ABP_mode}). As a first
application of (\ref{DK_ABP_mode}) we carry-on a linear
stability analysis of this unpolarized state, which will give
us a better understanding of the role of activity in the
structural transition. To this end we add perturbations,
$\rho_n(\bx,t)=\bar\rho\delta_{n0}+\delta\rho_n(\bx)\exp(\lambda t)$, inject this into (\ref{DK_ABP_mode}), and
linearize in the amplitudes $\delta\rho_n$. The convolution
products for $n \ne 0$ can be neglected since they are of second order
 in $\delta\rho_n$. The resulting
infinite set of equations is then truncated at an arbitrarily
large maximum value of $|n|$, say $M$ (i.e.
$\delta\rho_n=0~~\forall |n|>M$). It turns out that the
truncated set can be solved exactly for any $M$ giving an
equation for the growth rate $\lambda$. More explicitly, the set of
equations obtained by truncation to an arbitrary order $M$ is:
\BA
\widehat{\delta \rho_0} &= U_0 i \left(q_x \delta\widehat{\rho_1^R} + q_y \widehat{\delta\rho_1^I} \right)/ (\lambda + q^2 D + \bar\rho q^2 \tilde{V}(q)) ,   \nonumber \\
\widehat{\delta \rho_n^R} &= U_0 i \left[q_x (\widehat{\delta \rho_{n+1}^R}+\widehat{\delta \rho_{n-1}^R})+ q_y (\widehat{\delta \rho_{n+1}^I}-\widehat{\delta \rho_{n-1}^I}) \right] / c_n , \nonumber \\
\widehat{\delta \rho_n^I} &= U_0 i \left(q_x (\widehat{\delta \rho_{n+1}^I}+\widehat{\delta \rho_{n-1}^I})- q_y (\widehat{\delta \rho_{n+1}^R}-\widehat{\delta \rho_{n-1}^R}) \right) / c_n ,\nonumber \\
\widehat{\delta \rho_M^R} &= U_0 i \left(q_x \widehat{\delta\rho_{M-1}^R} - q_y \widehat{\delta\rho_{M-1}^I} \right) / c_M ,\nonumber \\
\widehat{\delta \rho_M^I} &= U_0 i \left(q_x
\widehat{\delta\rho_{M-1}^I} + q_y \widehat{\delta\rho_{M-1}^R}
\right) / c_M ,
 \label{SM:modesys}
\EA
where $1 \le n \le M-1$ and $\widehat{\delta\rho_n^R}$ and
$\widehat{\delta\rho_n^I}$ are respectively the spatial Fourier
transforms of the real and imaginary parts of the mode
amplitude $\delta\rho_n$: $\widehat{\delta \rho_n^R}(\bq) 
= \int \exp(-i \bq \cdot \bx)\mbox{Re}[\delta\rho_n] d\bx$ and 
$\widehat{\delta \rho_n^I}(\bq) = \int \exp(-i \bq \cdot \bx)\mbox{Im}[\delta\rho_n] d\bx$.
 $\widehat{\delta\rho_0}$ is the
spatial Fourier transform of $\delta\rho_0$. The coefficients
$c_n$ are given for any value of $n$ (including $M$) by
\BA
c_n = 2(\lambda + n^2 D_r + q^2 D) \ .
\EA

$\widehat{\delta\rho_M^R}$ and $\widehat{\delta\rho_M^I}$
depend only on corresponding quantities for the previous mode
$M-1$. If we inject the equations for mode $M$ into those for
$M-1$ and iterate this procedure for decreasing values of $n$,
we can write $\widehat{\delta \rho_n^R}$ and $\widehat{\delta
\rho_n^I}$, $n=M,M-1,...,1$ in terms of
$\widehat{\delta\rho_0}$. Using this into the first equation
of~(\ref{SM:modesys}), we get a polynomial equation of order
$M$ for the growth rate $\lambda$:
\BE
\lambda + q^2 \left[ D + U_0^2 f(\lambda,q) \right] + \bar\rho
q^2 \tilde{V}(q) = 0 \ .
\label{lambda_eq}
\EE
The function $f(\lambda,q)$ depends on the parameters $D$,
$D_r$ and $U_0$, but not on $\bar\rho$ nor on the interaction
potential. It is given by a continuous fraction:
\BE
f(\lambda,q) = \frac{1}{d_1 + \frac{1}{d_2 + \frac{1}{\ldots
+ \frac{1}{d_{M-1}+\frac{1}{d_M}}}}}
\label{SM:flambda}
\EE
with
\BE
d_n = \left\{
\begin{array}{ll}
c_n &\mbox{if $n$ is odd,}\\
c_n/U_0^2 q^2 &\mbox{if $n$ is even.}
\end{array}
\right.
\EE
Therefore, (\ref{lambda_eq}) is a polynomial equation of order
$M+1$ for $\lambda$. We thus have $M+1$ solutions for $\lambda$.
When $D_r \to \infty$ or $U_0 \to 0$ we have
$\mbox{Re}[\lambda]\to-\infty$ for all solutions except one branch
which converges to the expression valid for passive particles
\cite{Delfau2016}: $\lambda=-q^2(D+\bar\rho\tilde{V}(q))$. For
active particles the passive diffusion coefficient $D$ is
replaced by a frequency- and wavenumber-dependent generalized
diffusion coefficient $D(\lambda,q)\equiv D + U_0^2
f(\lambda,q)$, showing that the angular modes add memory and
non-locality to the linearized dynamics close to the
homogeneous-isotropic state. When $D_r$ is large or at large
scales ($q\to0)$ we find for the largest growth rate:
$\lambda\approx -q^2(D_{eff}+\bar\rho\tilde{V}(q))$ with an
effective diffusion coefficient $D_{eff} = D + U_0^2 /(2 D_r)$.
The idea of self-propulsion being equivalent to an enhanced
diffusion given by this $D_{eff}$ has been pointed out by
several authors in the dilute limit
\cite{Loi2008,Palacci2010,Fily2012}, but we have not made any
assumption regarding the density of the system so that our
expressions remain valid at high densities. As
mentioned, the Brownian dynamics is obtained for very large
$D_r$, or rather very small $U_0$. The behavior for $q\to 0$
guarantees that for repulsive $V(\bx)$ (for which $\tilde
V(q=0)>0$) there is no long-wavelength
spinodal-decomposition-like instability as in the case of MIPS.
Instead we can only have finite-wavelength instabilities. 
Note that our expressions for the instability of the homogeneous 
state can be used to elucidate if particular potentials 
beyond the GEM class explicitly discussed here lead to crystal 
formation, as long as the potential is sufficiently soft to 
justify the mean-field approximation which leads to Eq.~\ref{DK_ABP}.

\begin{figure}[ht!]
\includegraphics[width=\columnwidth]{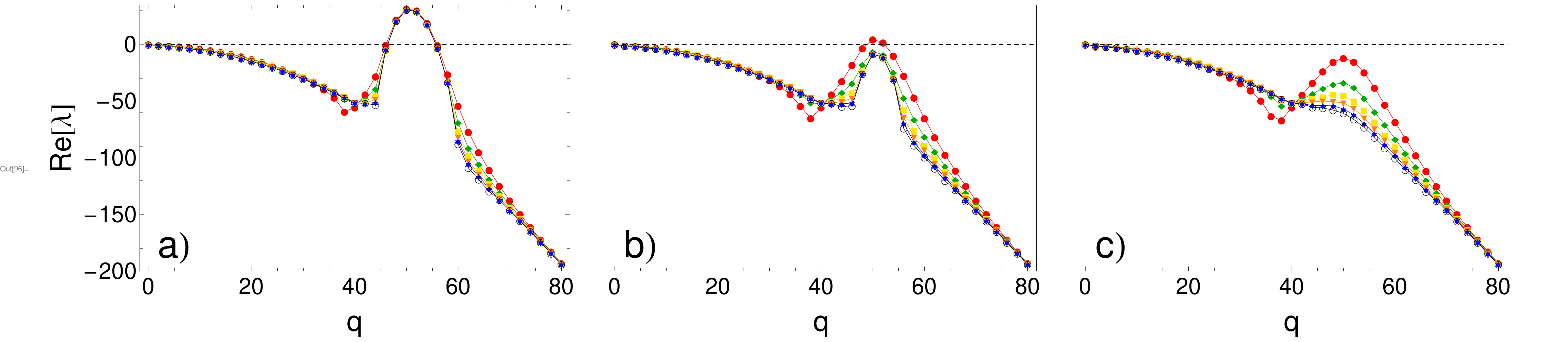}
\caption{Largest branch of the real part of the growth rate given by solution of (\ref{lambda_eq}) for $\bar\rho = 2000$,
$\epsilon = 0.0333$, $D_r = 0.1$, $R = 0.1$ and $D = 3 \times 10^{-2}$. The number of modes $M$ is 3 (red disks),
5 (green diamonds), 7 (yellow squares), 9 (orange triangles), 11 (blue stars) and 13 (black circles). a) $U_0 = 1.75$,
b) $U_0 = 2.4$ and c) $U_0 = 3$.}
\label{fig:SMdispersion_modes}
\end{figure}

We can now check the accuracy of truncating at different values
of $M$. If enough modes $M$ have been included, the predictions
obtained for higher values of $M$ should collapse on the same
curve. It appears that the higher the self-propulsion, the
higher $M$ needs to be to reach convergence: indeed,
figure~\ref{fig:SMdispersion_modes} shows that $M=3$ modes
 are enough for $U_0=1.75$, but we need at least $M=5$ modes
for $U_0=2.4$. Note that for $U_0 = 3$, $M=13$ modes are barely
enough. Therefore, keeping only a small number of modes
 will accurately describe the structural transition only if this one takes
place for a small value of $U_0$, meaning that the system is
already quite close to the critical point. However, we will see in the following
 subsection that they contain qualitatively the relevant mechanisms
shaping the cluster crystals.

\begin{figure}  [ht!]
\centerline{\includegraphics[width=.8\columnwidth]{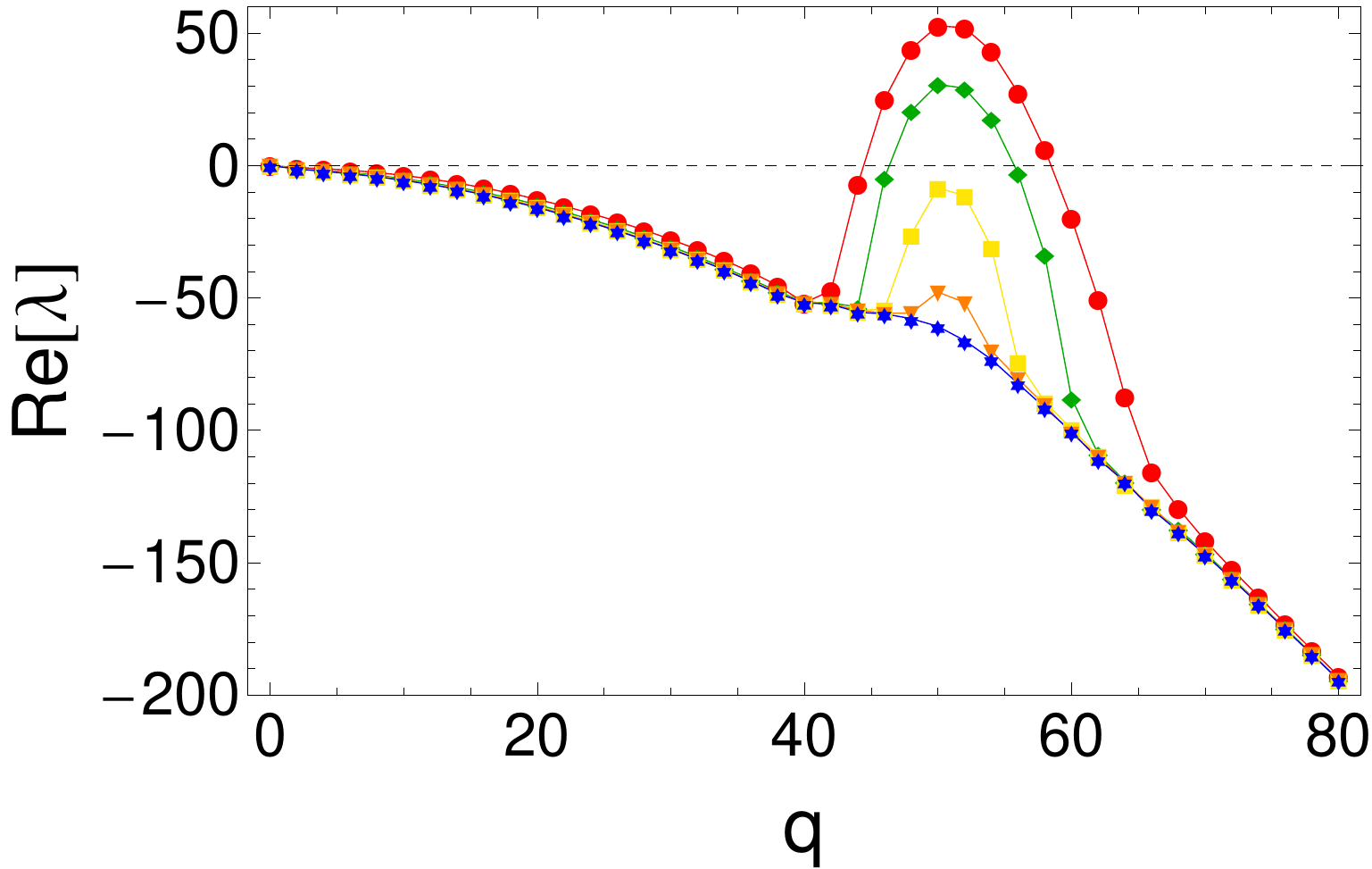}}
\caption{Largest branch of the real part of the growth rate $\lambda$ given by solution of
(\ref{lambda_eq}). Same parameters as in
figure~\ref{fig:snap_ldensity} a) except $U_0 = 1$ (red disks),
$1.75$ (green diamonds), $2.4$ (yellow squares),
$2.7$ (orange triangles) and $3$ (blue stars). Continuous fraction defining $f(\lambda,q)$ truncated
at $M=13$.}
\label{fig:dispersion_U0}
\end{figure}

Figure~\ref{fig:dispersion_U0} shows the maximum real part of
the growth rate as a function of $q$ for several values of
$U_0$. The continuous fraction defining $f(\lambda,q)$ has been
truncated to $M=13$, which is sufficiently large for the
results to remain unchanged when increasing $M$ further.
In the passive case \cite{Delfau2016} and for
the GEM-3 potential the homogeneous state becomes first
unstable with respect to a wavenumber $q_c$ depending only on
the interaction range $R$ and satisfying $q_c R\approx 5.0$.
This corresponds to a cluster crystal of periodicity $2\pi/q_c
\approx 1.26 R$. For the parameters used here, for which
$R=0.1$, we have $q_c\approx 50$. In
Fig.~\ref{fig:dispersion_U0} we see that a very similar
wavenumber is the most unstable in the active case for small
$U_0$. When $U_0$ is sufficiently increased though, these
positive growth rates eventually become negative. Activity thus
stabilizes the homogeneous state and prevents clustering, in
agreement with our particle simulations.

More quantitatively, one can compare the values of the
transition threshold $U_0^c$ obtained analytically for
sufficiently large $M$, and from particle simulations.
 For the second case we start from an initial
hexagonal crystal of clusters and increase $U_0$ step-by-step
until we get a long-lived gas phase. Table~\ref{SM:table1}
shows that the linear stability analysis gives us the right
order of magnitude but it systematically underestimates the
threshold. We attribute this to the nature of the transition,
which in the passive-particle case is
subcritical and subject to hysteresis \cite{Delfau2016}. We
expect this subcritical nature to remain also for small $U_0$.
Also, the use of a deterministic version of the DK equation, as
an approximation to the complete stochastic one, may be a
source of error.

\begin{table}[ht!]
\vspace{0.5cm}
\begin{center}
\begin{tabular}{|l|c|c|c|c|c|}
\hline
\hline											
$10^2 \times D$    & 3.0     &4.0  	 & 5.0 & 5.4  \\
\hline
$U_0^c$ (simulations)   & 2.65  &   2.05 		& 1.45 & 1.35  \\
$U_0^c$ (linear theory)        & 2.33  & 	1.87    & 1.15 & 0.5  \\
\hline
\hline
\end{tabular}
\end{center}
\caption{Critical self-propulsion $U_0^c$ above which the
cluster crystal disappears, obtained from particle simulations
and from the linear stability analysis of
(\ref{DK_ABP_mode}). $\bar\rho = 2000$, $D_r = 0.1$ and a
GEM-3 potential with $R=0.1$ and $\epsilon=0.0333$.
\label{SM:table1}}
\end{table}

\subsection{Macroscopic equation for 2 modes}
\label{2modes_eq}

Equations (\ref{DK_ABP}) or (\ref{DK_ABP_mode}) are quite
involved and it would be desirable to work with a simpler set
of equations such as a small-$M$ truncation of
(\ref{DK_ABP_mode}). The linear analysis reveals that this is
in general not an accurate approach, as values up to $M=13$
need to be considered to have convergence of the growth rates.
This is so because, except for the largest values of $D_r$ or
the smallest $U_0$, the average velocity field at each point
has a well-defined direction (see for example figure
\ref{fig:snap_ldensity} b) and many angular modes are needed to
represent such localized distribution in $\theta$ space.
Nevertheless a truncation to order $M=1$ leads to relatively
simple equations which give insight into the physical
mechanisms and favor qualitative understanding. Following
Bertin et al. \cite{Bertin2009}, we consider a hypothetical
situation in which $\varepsilon=|\rho_1/\rho_0|<<1$ and then
$\rho_n(\bx,t) = \mathcal{O}(\varepsilon^{|n|})$, so that we
can neglect the modes beyond $n=\pm 1$ (truncation to $M=1$).
We get:
\BA
\partial_t \rho(\bx,t) &= -\nabla \cdot \left( \bfr(\bx,t) \rho(\bx,t) \right)
+ D \nabla^2 \rho(\bx,t) - U_0 \nabla \cdot \bP(\bx,t) \nonumber\\
\partial_t \bP(\bx,t) &= -\nabla \cdot \left( \bfr(\bx,t) \bP(\bx,t) \right)
+ D \nabla^2 \bP(\bx,t) - D_r \bP(\bx,t)  \nonumber\\
& - \frac{U_0}{2} \nabla \rho(\bx,t)\ , \label{DK_ABP_fin}
\EA
where $\bP(\bx,t) = \left(\begin{array}{c}
\mbox{Re}\left[\rho_1\right] \\
\mbox{Im}\left[\rho_1\right]
\end{array} \right) = \int \rho(\bx,\theta,t) {\bf \hat n}(\theta) d \theta$
is the momentum or polarization field. The product $\bfr\bP$
appearing in (\ref{DK_ABP_fin}) is a tensor product. Even if
Eqs.~(\ref{DK_ABP_fin}) are a rough approximation to
(\ref{DK_ABP_mode}) they still describe the system behavior
qualitatively. Integrating numerically
Eqs.~(\ref{DK_ABP_fin}) with a spectral method using $512
\times 512$ grid points, we recover the crystal of clusters
for small values of $U_0$ (see figure~\ref{fig:snap_ldensity}
c) and a homogeneous state for higher $U_0$. The polarization
structure of the clusters is also in good agreement with our
previous observations: Fig.~\ref{fig:snap_ldensity} d) shows
that the polarization field inside the clusters is radial.
Empty clusters are also observed in one dimension for small $D$
and high enough $U_0$. Having to work at small values of $D$
makes it more difficult to obtain empty clusters in two
dimensions because of the need of a high numerical resolution.
We have shown however than Eqs.~(\ref{DK_ABP_fin}) support
clusters with a depletion in their center (see
figure~\ref{fig:clust_structure} b) if we replace the
convolution product defining $\boldsymbol{F}_{\rho}$ by an
effective confinement potential justified by the approximation
described in the next paragraph.

Eqs.~(\ref{DK_ABP_fin}) serve also as a starting point to
better understand  the mechanisms shaping cluster structure. As
in the passive case \cite{Delfau2016} we can focus on the case
of small $D$ so that a first approximation for the steady
crystal density is a set of delta functions at the lattice
points $\{\ba_i\}$: $\rho(\bx)\approx N_p \sum_{\{\ba_i\}}
\delta(\bx-\ba_i)$. The number of particles per cluster, $N_p$,
can be expressed in terms of $\bar\rho$ and the intercluster
distance $a$: $N_p=\bar\rho a^2 \sqrt{3}/2$. With this
approximation $\bfr(\bx) \approx - N_p \sum_{\{\ba_i\}} \nabla
V(\bx-\ba_i)$. To consider the structure of a narrow cluster
centered at $\bx={\bf 0}$ we keep in the lattice sum only the
central and the six neighboring clusters, and expand around
$\bx\approx {\bf 0}$. For a GEM-$\alpha$ potential with
$\alpha<2$ the dominant term is the interparticle repulsion
within the central cluster, so that $\bfr(\bx)$ points outwards
and, as it is observed, the aggregate disappears. But when
$\alpha>2$ the repulsion from the neighboring clusters prevails
and produces a confining effective force at $\bx\approx {\bf
0}$ overcoming the local repulsion. At first order in the
distance $\bx$ to the cluster center, we have $\bfr(\bx) \approx
-\gamma~ \bx$. $\gamma$ depends on $N_p$, on
the intercluster distance $a$ and on the interaction potential
parameters $\epsilon$ and $\rho$. Its precise expression is not
particularly illuminating, but it can be found as
$\gamma=D/\sigma_{2D}^2$ with $\sigma_{2D}$ given by Eq. (33)
in \cite{Delfau2016}. The harmonic character and radial
symmetry of this approximation reduces the problem in
Eqs.~(\ref{DK_ABP_fin}) to a stationary linear one for
$\rho(\bx)=\rho(r)$ and $\bP(\bx)=p_r(r){\bf\hat e_r}$ in polar
coordinates centered at the cluster center (${\bf\hat e_r}$ is
the unit vector in the radial direction) in the confining
harmonic force $-\gamma r$ arising from the repulsion by
neighboring clusters:
\BA
\label{radialeqs}
& D\rho'+\gamma r \rho -U_0 p_r = 0 &  \\
& D p_r'' + (\gamma r +\frac{D}{r}) p_r' + (2\gamma -D_r
-\frac{D}{r^2})p_r -\frac{U_0}{2} \rho' = 0  \ . & \nonumber
\EA

To get the equation for $\rho$, a first integral has already
been performed under the condition of zero net
particle flux in or out of the cluster, as appropriate for the
steady state. The system (\ref{radialeqs})
should be solved in $r\in[0,\infty]$, but the approximations
used are valid only if it gives a cluster width much smaller
than the interaction range $R$. The first equation in
(\ref{radialeqs}) shows that in the limit of small $U_0$, this
cluster width is of the order of $w \approx \sqrt{D/\gamma}$.
Eqs. (\ref{radialeqs}) require three initial or boundary
conditions, which could be taken as the values of $\rho$, $p_r$
and $p_r'$ at $r=0$. Regularity of the field $\bP(\bx)$ at the
origin implies $p_r(0)=0$ --- which was observed for
chemorepulsive active colloids \cite{Liebchen2015}--- and
$p_r(r\approx 0)\approx \nu r$ (so that $\nu=p_r'(0)$).
$\rho(0)$ can be determined by fixing the number of particles
in the cluster $\int_0^\infty 2\pi r dr \rho(r)=N_p$. In terms
of it and of $p_r(r)$ the first equation in (\ref{radialeqs})
can be solved explicitly:
\BE
\rho(r)=\left[ \rho(0)+\frac{U_0}{D}\int_0^r  e^{\frac{\gamma
u^2}{2 D}} p_r(u) du \right]e^{-\frac{\gamma r^2}{2 D}},
\EE
which shows that there is a shape change, in agreement with the
particle simulations, from a maximum of density at the origin
($\rho''(0)<0$) to a minimum ($\rho''(0)>0$) when the
polarization slope at the origin $\nu=p_r'(0)$ changes from
$\nu U_0 < \gamma \rho(0)$ to $\nu U_0 > \gamma \rho(0)$,
respectively. Finally, the condition $|\bP(\bx)| \le \rho(\bx)$
--- resulting from the definition of $\bP$ --- imposes the
value of this slope: $\nu$ must indeed cancel the prefactor of
the slow decay at large $r$, $p_r\sim r^{-\mu}$ with
$\mu=2-(D_r/\gamma)-U_0^2/(2 D\gamma)$, arising from the
large-$r$ behavior of the hypergeometric function that solves
the homogeneous part of the (\ref{radialeqs}) for $p_r(r)$.
This determination can only be done numerically.

\begin{figure}  
\centerline{\includegraphics[width=.8\columnwidth]{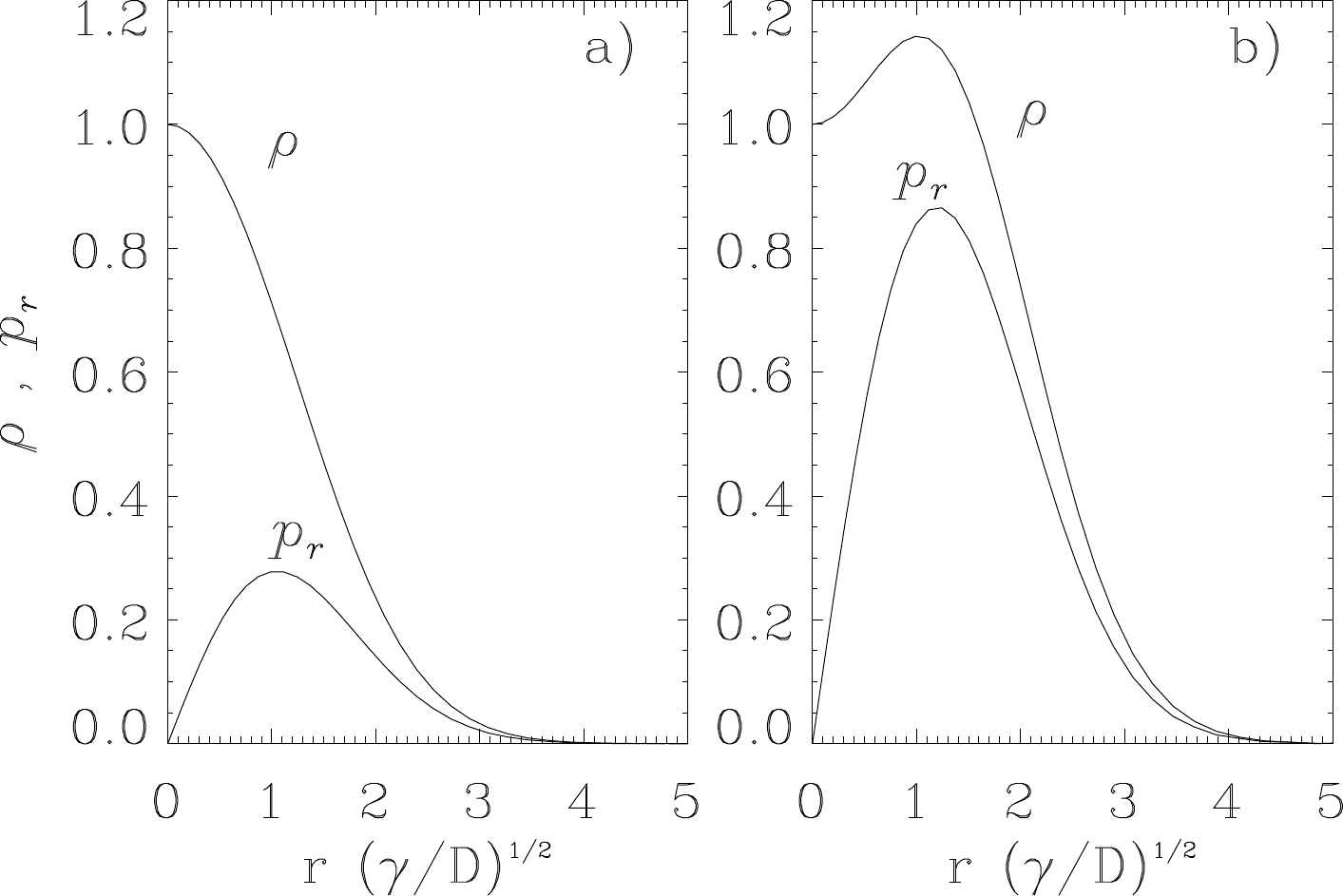}}
\caption{Particle density $\rho(r)$ and radial component of the polarization $p_r(r)$ from Eqs.
(\ref{radialeqs}). He have taken $\rho(0)=1$ and a dimensionless formulation in which the parameters are
$\tilde D_r=D_r/\gamma=10^{-6}$ and $\tilde U_0=U_0/\sqrt{\gamma D}$. a) $\tilde U_0=0.8$.
b) $\tilde U_0=1.4$. }
\label{fig:radial}
\end{figure}

Figure~\ref{fig:radial} shows examples of solutions of the
radial equations (\ref{radialeqs}), displaying the two
qualitatively different cluster shapes, with smaller/larger
density at the center. Although the truncation to $M=1$ leading
to Eqs. (\ref{DK_ABP_fin}) and (\ref{radialeqs}) precludes
quantitative agreement with particle simulations,
Fig.~\ref{fig:radial} (see also
figure~\ref{fig:clust_structure} b) shows that this 2-mode
truncation contains the essential qualitative mechanisms needed
for the cluster-crystal formation and the structure of the
clusters: an effective potential which confines  particles
within clusters, originating from the repulsion by the
neighboring clusters, and the presence of a polarization field
$\bP$, increasing with $U_0$, which pushes the particles
towards the periphery of the clusters until destroying them.

\section{Conclusion}

We have shown that ACCs occur in
systems of active Brownian particles with soft repulsive interactions. Self-propulsion
deforms the  clusters by depleting particle
density inside, and large self-propulsion stabilizes the
homogenous-isotropic state. We have derived a
continuous description and analyzed
the crystal forming instability by linear analysis. Truncation
to two angular modes, despite not being quantitatively
accurate, retains the basic mechanisms. In particular it allows
to understand crystal persistence as an effect of the confining
forces arising from repulsion by neighboring clusters, and the
cluster shape as a balance between the confining force and the
tendency to radial escape driven by the polarization field.\\

We acknowledge financial support from grants LAOP, CTM2015-66407-P
(AEI/FEDER, EU) and ESOTECOS, FIS2015-63628-C2-1-R (AEI/FEDER, EU).

\clearpage
\section*{References}

\bibliographystyle{apsrev}
\bibliography{biblio}

\end{document}